\documentclass[a4paper,11pt]{article}
\pdfoutput=1 

\usepackage{jheppub} 

\usepackage{subcaption}

\newcommand{\hc}{\ensuremath{\mathrm{h.c.}}}
\newcommand{\braket}[1]{\ensuremath{\left\langle#1\right\rangle}}
\newcommand{\Eqref}[1]{Eq.~(\ref{#1})}

\title{Sterile Neutrino Dark Matter Production in the Neutrino-phillic Two Higgs Doublet Model}

\author[a]{Adisorn Adulpravitchai}
\author[b]{and Michael A.~Schmidt}

\affiliation[a]{Department of Physics, Faculty of Science, \\Chulalongkorn University,  Bangkok 10330, Thailand}
\affiliation[b]{ARC Centre of Excellence for Particle Physics at the Terascale,\\
School of Physics, The University of Sydney, NSW 2006, Australia}

\emailAdd{adisorn.adulpravitchai@gmail.com}
\emailAdd{michael.schmidt@sydney.edu.au}

\abstract{Sterile Neutrinos with a mass in the keV range form a good candidate for dark matter. They are naturally produced from neutrino oscillations via their mixing with the active neutrinos. However the production via non-resonant neutrino oscillations has recently been ruled out. The alternative production via Higgs decay is negligibly small compared to neutrino oscillations. We show that in the neutrino-phillic two Higgs doublet model, the contribution from Higgs decay can dominate over the contribution from neutrino oscillations and evade all constraints. We also study the free-streaming horizon and find that a sterile neutrino mass in the range of 4 to 53 keV leads to warm dark matter.}

\begin{document}
\maketitle

\section{Introduction}

The Standard Model (SM) of particle physics is very successful, but it fails to explain neutrino mass and dark matter (DM). Dark matter accounts for about one quarter of the energy density of the Universe, five times more than ordinary matter, but its origin is unknown. A good candidate for the dark matter are sterile neutrinos with a keV-scale mass and tiny mixing with the active neutrinos, which is a simple extension of the SM~\cite{Abazajian:2012ys,Drewes:2013gca}. In contrast to standard cold dark matter, they are generally warmer with a larger free-streaming horizon. Thus they are candidates for warm dark matter and suppress structure at small scales addressing the missing satellite problem~\cite{Kauffmann:1993gv,Klypin:1999uc,Moore:2005jj} and possibly explaining the velocities of pulsars~\cite{Kusenko:2006rh,Kusenko:1997sp}.

There are many ways of producing sterile neutrinos: (i) they can be produced through neutrinos oscillations in the early Universe via a small mixing with the active neutrinos~\cite{Barbieri:1990vx,Enqvist:1990ad,Dodelson:1993je}. This mechanism is already excluded by observation~\cite{Horiuchi:2013noa}, but the bounds can be avoided, if there is a large enough primordial lepton asymmetry and sterile neutrinos are produced via resonant oscillations~\cite{Shi:1998km}. (ii) Another well-studied alternative is non-thermal production via decay of a scalar field in thermal equilibrium~\cite{Shaposhnikov:2006xi,Bezrukov:2009yw,Kusenko:2006rh,Petraki:2007gq,Frigerio:2014ifa,Merle:2015oja}, or a scalar produced via the freeze-in mechanism~\cite{Hall:2009bx}, which subsequently decays to sterile neutrinos~\cite{Merle:2013wta,Adulpravitchai:2014xna,Kang:2014cia,Merle:2015oja}. Recently several alternative production mechanisms from decay have been suggested such as the production from the decay of pions~\cite{Lello:2014yha}, Dirac fermions~\cite{Abada:2014zra}, light vector bosons~\cite{Shuve:2014doa}, or a condensate formed during inflation~\cite{Enqvist:2014zqa,Nurmi:2015ema}. (iii) Finally the keV sterile neutrinos could have been in thermal equilibrium and their abundance diluted by production of entropy~\cite{Bezrukov:2009th,Nemevsek:2012cd,Bezrukov:2012as,Tsuyuki:2014aia,Patwardhan:2015kga}. 

In any model with mixing between active neutrinos and the keV sterile neutrinos a fraction of the sterile neutrino abundance will be generated via neutrino oscillations. The mixing is generally induced after an electroweak doublet scalar obtains a vacuum expectation value (vev). The Yukawa interaction however induces a second contribution: The electroweak doublet scalar can decay into a SM lepton and a sterile neutrino. In a model with one Higgs doublet, this contribution is always negligible compared to the contribution from neutrino oscillations~\cite{Matsui:2015maa}, because the vev, $v=174$ GeV, and thus the mixing is sizeable. However this does not hold anymore in models with multiple Higgs doublets. The vev of one of the Higgs doublets might be tiny, smaller than $\mathcal{O}($MeV$)$, and consequently the production via Higgs decay might dominate. Sterile neutrino dark matter with a keV-scale mass has been previously considered in a two Higgs doublet model in Ref.~\cite{Haba:2014taa} and the production of the required number density via the decay of an electroweak doublet has been studied in Ref.~\cite{Molinaro:2014lfa} in the scotogenic model of neutrino mass~\cite{Ma:2006km}. 

We  consider a two Higgs doublet model, where one of the electroweak doublet scalars exclusively couples to the sterile neutrino and study the production of keV sterile neutrino DM via the decay of this electroweak doublet in detail. The main result is the momentum distribution function for the sterile neutrino DM and the free-streaming horizon, which we use to determine the relevant parameter space where the keV sterile neutrino constitutes warm dark matter. This mechanism can be easily embedded in a seesaw~\cite{Ma:2000cc,Haba:2014taa} or radiative~\cite{Ma:2006km,Molinaro:2014lfa} neutrino mass model.

The paper is organised as follows: In section \ref{section-Model}, we introduce the two-Higgs doublet model and discuss the mass spectrum. The produced sterile neutrino DM abundance is discussed in  section \ref{sec:DMprod}. Section \ref{sec:rFS} is dedicated to the free-streaming horizon of the sterile neutrinos and we briefly comment on the effective number of relativistic degrees of freedom in section \ref{sec:Neff}. In section \ref{sec:higgsed}, we discuss the possibility that the scalar doublet obtains a tiny vev, thus the sterile neutrinos can decay and explain the observed an X-ray line at 3.55 keV~\cite{Bulbul:2014sua,Boyarsky:2014jta}. We conclude in section \ref{sec:conclusions}. Technical details are collected in the appendices.

\section{Neutrino-phillic Two Higgs Doublet Model} \label{section-Model}
We consider a two Higgs doublet model  with a second scalar doublet $H_\nu$ which exclusively couples to the sterile neutrino $N$ and the left-handed lepton doublet $L$. This is guaranteed by a $Z_2$ symmetry under which the new fields, $N$ and $H_\nu$, are odd, but all SM particles are even. A coupling to other fermions is strongly constrained by flavour-violating processes. The most general Yukawa interactions in the lepton sector are given by 
\begin{equation}
-\mathcal{L}= y_{E} L H E^C + y_{LN} L H_\nu N    + \frac12 m_N N^2+\hc\;.
\end{equation}
and the most general scalar potential is defined in the usual way 
\begin{eqnarray}
V&=& -m_1^2 H^\dagger H +\frac{\lambda_1}{2} (H^\dagger H)^2+m_2^2 H_\nu^{\dagger} H_\nu +\frac{\lambda_2}{2} (H_\nu^{\dagger} H_\nu)^2 \\ \nonumber 
& &  +\lambda_3 H^\dagger H\, H_\nu^\dagger H_\nu +\lambda_4 |H^\dagger H_\nu|^2+\frac{\lambda_5}{2} [(H^\dagger H_\nu)^2 +\hc]\;,
\end{eqnarray}
where we used a field redefinition of $H_\nu$ to absorb the complex phase of $\lambda_5$. The Yukawa couplings $y_{LN,\alpha}$ are generally complex.
After the SM Higgs doublet obtains a vev, $v^2=m_1^2/\lambda_1$, electroweak symmetry is broken and we decompose the fields in terms of their components
\begin{align}
H&=\begin{pmatrix}G^+\\ v+\frac{1}{\sqrt{2}} \left(h+ i G^0\right)\end{pmatrix}&
H_\nu&=\begin{pmatrix}K^+\\ \frac{1}{\sqrt{2}} \left(k+ i K^0\right)\end{pmatrix}\;.
\end{align}
The scalar masses at leading order are given by
\begin{align}\label{masses}
m_h^2 &= 2 m_1^2 = 2 \lambda_1 v^2  &
m_k^2 &= m_2^2+\left(\lambda_3+\lambda_4+\lambda_5 \right)v^2 \\\nonumber
m_{K^0}^2&= m_2^2+\left(\lambda_3+\lambda_4-\lambda_5\right) v^2 & 
m_{K^\pm}^2 &= m_2^2+\lambda_3 v^2\;,
\end{align}
where $h$ describes the observed Higgs boson at $m_h=125$ GeV~\cite{ATLAS:2012gk,CMS:2012gu}. As long as $H_\nu$ does not obtain a vev, there is no mixing between the different states. We will comment on this possibility in section \ref{sec:higgsed}.
The active neutrinos obtain mass in the usual way. Both Dirac and Majorana mass terms are possible. For example, one can introduce the Weinberg operator\cite{Weinberg:1979sa}, $LLHH$, to generate the neutrino mass. We will not discuss it further, because it does not affect the production of the keV sterile neutrino $N$.

\section{Dark Matter Production}\label{sec:DMprod}
Freeze-in production of sterile neutrino DM with a second Higgs doublet has been studied in the scotogenic model~\cite{Molinaro:2014lfa}. Here we focus on keV sterile neutrinos and do not specify the mechanism of neutrino mass generation explicitly.

The keV sterile neutrino can be produced via the decay of the scalar fields $(k,K^0,K^{\pm})$ while they are in thermal equilibrium.
 We will present a crude simple calculation using the Maxwell-Boltzmann approximation and neglect Pauli-blocking. The result will result in a good order of magnitude estimate. In our discussion, however, we will use the more accurate result using the distribution function, which can be found in App.~\ref{app:TempSterileNu}.
Assuming that inverse decays can be neglected, the yield $Y(T)=n(T)/s(T)$ of the sterile neutrino can be obtained from the Boltzmann equation,
\begin{equation}
s T \frac{d Y_{N_1}}{d T} = -\frac{\gamma_{N_1}(T)}{H(T)},
\end{equation}  
where $s$ is the entropy density of the Universe, $H(T)$ is the Hubble parameter at a given temperature and $\gamma_{N_1}(T)$ is the thermally averaged sterile neutrino production rate,
\begin{equation}
\gamma_{N_1}(T) = \sum_X \frac{g_X m_X^2 T}{2 \pi^2} K_1(m_X/T) \Gamma(X\rightarrow N_1 l),
\end{equation}
where $X=k,K^0,K^{\pm}$ and $l$ is a SM lepton.
Following the freeze-in calculation in Ref.~\cite{Hall:2009bx}, we can integrate the equation and obtain the final yield of sterile neutrinos after freeze-in
\begin{equation}\label{eq:Yinf0}
Y^\infty_{N_1}=\frac{45 }{4 \pi^4  }\sum_X \frac{g_X\Gamma(X\to N_1 l)M_{0}}{m_X^2 g_*^s(T_d)}\int_{x_{min}}^{x_{max}}x^3K_1(x)dx
\end{equation}
where we defined the typical mass scale 
\begin{equation}\label{eq:M0}
M_0=\frac{3}{2\pi}\sqrt{\frac{5}{\pi g_*^{\rho}(T_d)}}M_{Pl}\;.
\end{equation}
This allows us to rewrite the Friedmann equation during the radiation dominated epoch as
\begin{equation}
H = \frac{T^2}{M_0}
\end{equation}
using the Planck mass $M_{Pl}$ and the usual definition of the effective entropy degrees of freedom $g_*^s$ and the relativistic degrees of freedom $g_*^\rho$
\begin{align}
s&=\frac{2\pi^2}{45}g_*^s(T) T^3&
\rho&=\frac{\pi^2}{30}g_*^\rho(T) T^4\;.
\end{align}
We can obtain an approximate analytic solution to this equation by extending the integration over all positive values of $x$, i.e. $x_{min}\to 0$ and $x_{max}\to \infty$
\begin{align}\label{eq:Yinf1}
Y^\infty_{N_1}&\simeq\frac{405\sqrt{5} }{16 \pi^{9/2}  }\sum_X \frac{g_X\Gamma(X\to N_1 l)M_{Pl}}{m_X^2g_*^s(T_{d,X})\sqrt{g_*^\rho(T_{d,X})}}\\
&\simeq 0.328 \sum_X \frac{g_X\Gamma(X\to N_1 l)M_{Pl}}{m_X^2g_*^s(T_{d,X})\sqrt{g_*^\rho(T_{d,X})}}\;.
\end{align}

In order to obtain simple analytic results, we made several (crude) approximations: 
(i) We used the Maxwell-Boltzmann approximation and thus also neglected Pauli-blocking of the neutrinos.
(ii) We extended the integration boundaries in \eqref{eq:Yinf0} to obtain the leading order result in Eq.~\eqref{eq:Yinf1}. This is justified by noting that freeze-in is typically dominated by processes around $T\sim m_X$\cite{Hall:2009bx}.
(iii) We assumed that the effective number of relativistic degrees of freedom for entropy, $g_*^s$, and energy, $g_*^\rho$, do not change during the production of dark matter, which is reasonably well satisfied for scalar masses $m_X\gtrsim 100$ GeV. 
(iv) We neglected $2\leftrightarrow2$ scattering processes like $X^0 + \ell^\pm  \to N+ W^\pm$ and $X^\pm + \nu \to N+ W^\pm$. These processes are subdominant compared to two body decays of $X$ due to phase space suppression. 
(v) We assumed that the particles $X$ are in thermal equilibrium until freeze-in occurs at $x_{fi}\sim2-5$~\cite{Hall:2009bx} and thus $T_{fi}\gtrsim 20$ GeV for $m_X\gtrsim 100$ GeV. For sufficiently large Higgs portal couplings\footnote{Note that the Higgs portal couplings enter the freeze-in calculation only indirectly via the mass $m_X$ of the particle $X$.}  $\lambda_{3,5}$, the scattering of the scalars $X$ with $b$ quarks, $X + b \to X+b$, will keep the scalars $X$ in thermal equilibrium similar to the SM Higgs down to $T\sim 5$ GeV.  
(vi) Finally we use the usual vacuum decay rate and do not take into account finite temperature effects, which has been studied e.g.~in Ref.~\cite{1305.0267v3}. We expect these corrections to be small, because the Yukawa coupling is small and the sterile neutrino a gauge singlet. 
In the following we will, however, use the more accurate result in Eq.~\eqref{eq:numDensity}. 
\begin{equation}
Y^\infty_{N_1}\simeq 0.207 \sum_X \frac{g_X\Gamma(X\to N_1 l)M_{Pl}}{m_X^2g_*^s(T_{d,X})\sqrt{g_*^\rho(T_{d,X})}}\;.
\end{equation}
It has been obtained by solving the Boltzmann equation for the distribution function without using the Maxwell-Boltzmann approximation and including Pauli-blocking. The detailed calculation is outlined in App.~\ref{app:distrib} and \ref{app:TempSterileNu}.
\begin{figure}[tb]\centering
\begin{subfigure}{0.48\linewidth}\centering
\includegraphics[width=\linewidth]{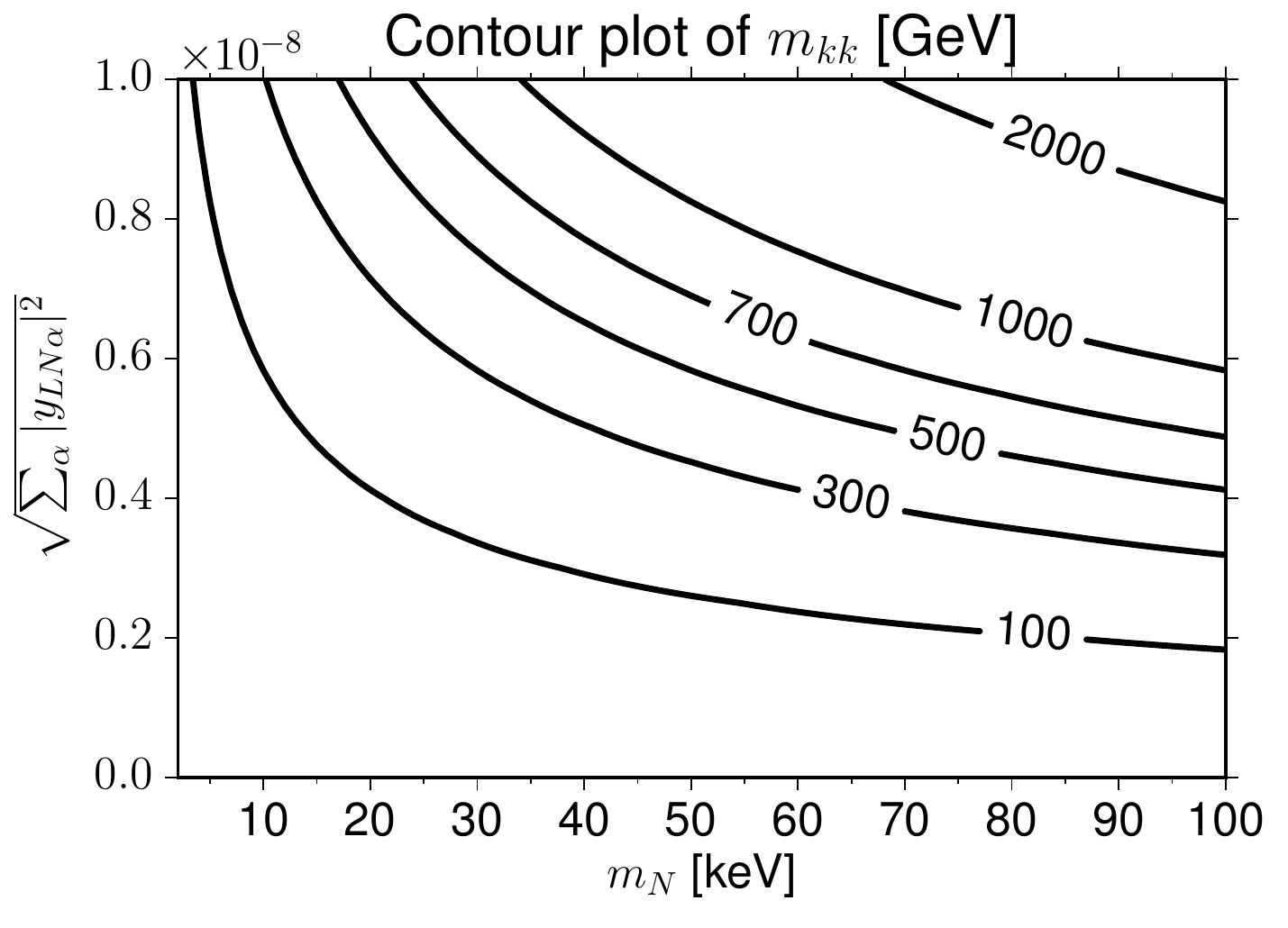}
\caption{$m_N$ vs.~$\sqrt{\sum_\alpha |y_{LN,\alpha}|^2}$. The contour labels are the scalar mass $m_{kk}$ in GeV.}
\label{AA1}
\end{subfigure}
\hfill
\begin{subfigure}{0.48\linewidth}\centering
\includegraphics[width=\linewidth]{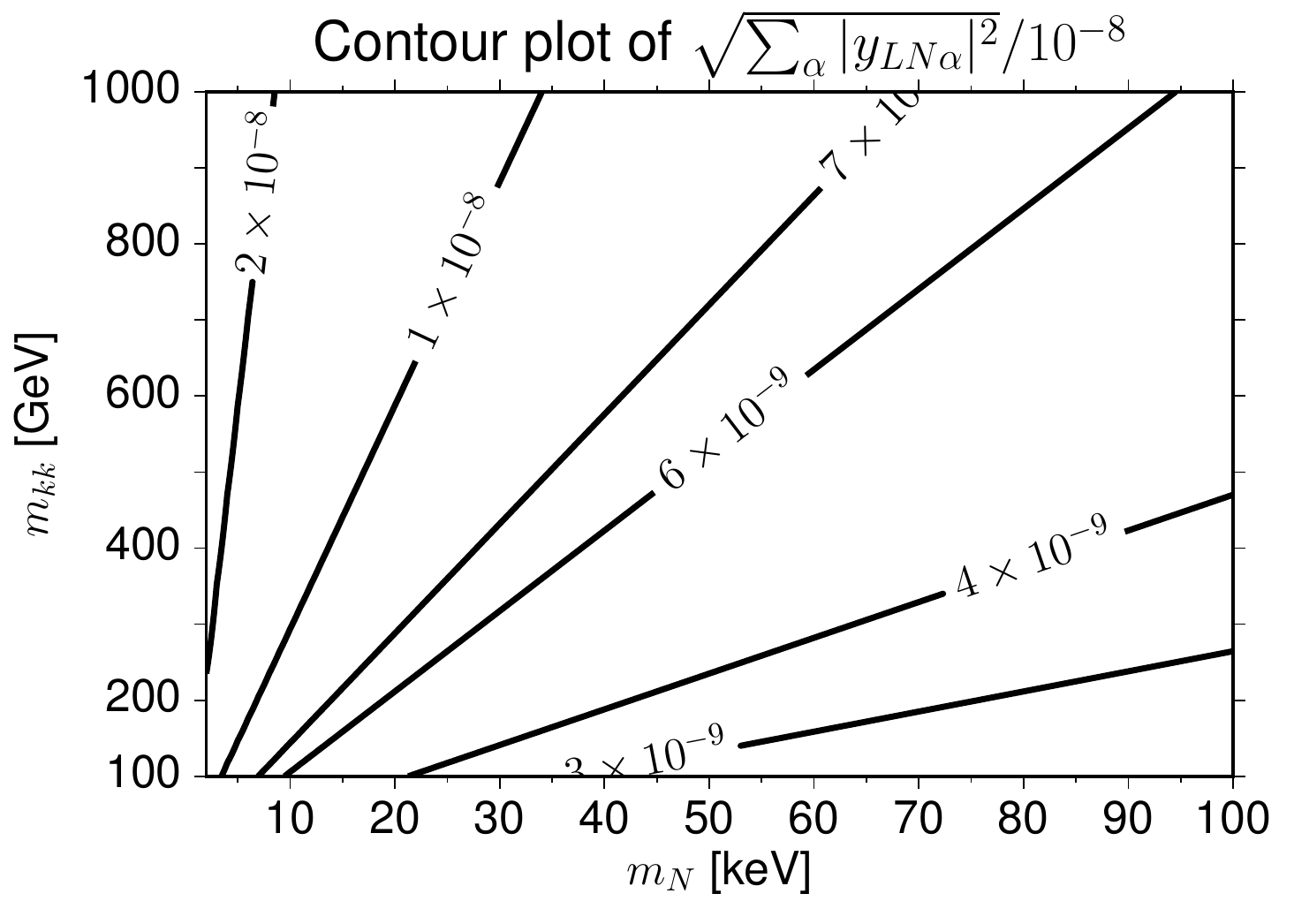}
\caption{$m_{kk}$ vs.~$m_N$. The contour labels are the effective coupling $\sqrt{\sum_\alpha |y_{LN,\alpha}|^2}$.}
\label{CC1}
\end{subfigure}
\caption{Contour plots with fixed DM relic abundance $\Omega_{N_1} h^2=0.1199$~\cite{Ade:2015xua}.}
\label{fig:DMprod}
\end{figure}

The number of degree of freedom of the scalar fields  are given by $g_k=1, g_{K^0}=1,g_{K^+}=2$ and the decay widths $\Gamma_X$ are given in App.~\ref{app:decaywidth}.
We finally obtain the relic abundance of the sterile neutrino dark matter using $\Omega_{N_1}=m_N s_0 Y_{N_1}^\infty/\rho_{cr}$ with the critical energy density $\rho_{cr}=3 H^2 M_{Pl}^2/8\pi$ and find 
\begin{equation}
\Omega_{N_1} h^2 \simeq 6.88 \times 10^{26} m_{N} \sum_X \frac{g_X\Gamma(X\to N_1 l)}{m_X^2 g_*^s(T_{d,X}) \sqrt{g_*^\rho(T_{d,X})}}\;.
\end{equation} 
Taking the limit of equal scalar masses, $m_k \simeq m_{K^0} \simeq m_{K^+}\equiv m_{kk}$ and $T_{d,X}\equiv T_d$, the relic abundance 
\begin{equation}
\Omega_{N_1} h^2 \simeq \frac{6.88 \times 10^{26}}{g_*^s(T_d) \sqrt{g_*^\rho(T_d)}}  \frac{3 m_{N}}{16 \pi m_{kk}} \sum_\alpha |y_{LN,\alpha}|^2
\end{equation} 
only depends on three parameters: the two masses $m_{kk}$, $m_N$ and the effective coupling $\sum_\alpha |y_{LN,\alpha}|^2$. Fixing the dark matter relic abundance to the observed best fit value for dark matter, $\Omega_{\mathrm{DM}} h^2=0.1199$, by Planck~\cite{Ade:2015xua}, we show in Fig.~\ref{fig:DMprod} two contour plots illustrating the dependence on the three parameters. We find that the effective coupling has to be of order $10^{-9}$ for keV sterile neutrino masses in the range $2-100$ keV and scalar doublets with electroweak-scale masses. Although there is no explanation for the smallness of the Yukawa coupling, it is technically natural. 
Larger scalar masses generally require smaller couplings to compensate for the suppression by the scalar mass $m_{kk}$. Larger sterile neutrino masses generally require larger effective couplings or larger scalar masses, because the DM abundance is proportional to the ratio $m_N/m_{kk}$.

\section{Free Streaming Horizon}\label{sec:rFS}
The free streaming horizon characterises the scale below which perturbations are suppressed in the power spectrum~\cite{Boyarsky:2008xj}. It is defined by the average distance a particle travels without any collisions
\begin{equation}\label{eq:FreeStreamingHorizon}
r_{\rm FS} = \int_{t_{in}}^{t_{0}} \frac{\langle v \rangle}{a(t)} dt
\end{equation}
where $\braket{v}$ is the average velocity, $t_{in}$  denotes the time when the sterile neutrino is produced and $t_0$ the time  today.
In order to evaluate it, we have to find the average velocity of the sterile neutrino
\begin{equation}
\langle v \rangle =  \braket{\frac{p}{E}} 
=\begin{cases}
 1  &{\rm for } \;\; t <  t_{N,nr}  \\ 
\frac{\braket{p}}{m_N} & {\rm for } \;\; t> t_{N,nr}
\end{cases}
\end{equation}
where $t_{nr}$ ($T_{N,nr}$)  denotes the time (temperature of the sterile neutrinos) when neutrinos become non-relativistic, i.e.
\begin{equation}
 \braket{p(T_{N,nr})} = m_N\;.
\end{equation}
We can define the average momentum in terms of the momentum distribution function\footnote{Note that we limit our discussion to a homogeneous isotropic Universe which is described by the Friedmann-Robertson-Walker metric and thus also take the distribution function to be homogeneous.}  $f(p,t)$ of the sterile neutrinos, which can be obtained from the Boltzmann equation
\begin{equation}
L[f]=C[f]
\end{equation}
with the Liouville operator $L[f]$ and the collision term $C[f]$. The Liouville operator is defined as~\cite{Kolb:1990vq}
\begin{equation}
L[f]=\left(\frac{\partial }{\partial t} -H p\frac{\partial}{\partial p}\right) f(p,t)\;.
\end{equation}
In analogy to the treatment in Refs.~\cite{Petraki:2007gq,Merle:2015oja}, we introduce the dimensionless quantity
\begin{align}\label{eq:dimParam}
x_N&=\frac{p_N}{T_N}\;,
\end{align}
which allows to write the distribution function as
\begin{equation}
f_N(x_N,T_N) = \frac{2\sqrt{90\pi}}{\pi^{2}} \sum_X\frac{g_X\Gamma_X}{m_X^2\sqrt{g_*^\rho(T_d,X)}}
  \int_0^{\frac{m_X^2}{8 x_N T_{N}^2}} d y \sqrt{\frac{y}{x_N}} g\left(e^{y+\frac{x_N}{2}}\right)
\end{equation}
with 
\begin{equation}
g(z)=-\frac{1}{1+z}-\frac12\ln\left(\frac{z-1}{z+1}\right)\;.
\end{equation}
The scalar particle mass (decay width) is denoted $m_X$ ($\Gamma_X$) and we assumed that the effective number of relativistic degrees of freedom, $g_*^\rho$, remains constant during production, $T_{d,X}$ is the decay temperature of particle $X$ and we neglected the back reaction from inverse decays.  See appendix App.~\ref{app:distrib} for a derivation of the distribution function.
As the integrand is exponentially suppressed for $y\gg 1$, we can take the upper limit of the integral to infinity for temperatures $T_N\ll m_X/\sqrt{8 x_N}$. This is generally justified at late times, when $p_N\sim T_N \ll m_X$, thus we obtain for $f_N^0(x_N) \equiv \lim_{T_N\to 0} f_N(x_N,T_N)$
\begin{align}
f_N^0(x_N) & = \frac{\sqrt{90}}{2\pi}  \sum_X\frac{g_X\Gamma_XM_{pl}}{m_X^2 \sqrt{g_*^\rho(T_{d,X})}}
 \frac{1}{\sqrt{x_N}} \left[ 
2 \mathrm{Li}_{\frac32}\left(-e^{-\frac{x_N}{2}}\right)
+\mathrm{Li}_{\frac52}\left(e^{-\frac{x_N}{2}}\right)
-\mathrm{Li}_{\frac52}\left(-e^{-\frac{x_N}{2}}\right)
\right]
\;.
\end{align}
In this approximation the function $x^2 f_N(x)$ has only one maximum at $\hat x\simeq 1.54$ and it falls off very quickly away from the maximum. A typical value for the momentum is thus of order $\hat x T_N$. This justifies the approximation for small temperatures.
The average momentum is given by
\begin{equation}
 \braket{p(T_N)} \simeq 2.46\, T_N \label{momentump} 
\end{equation}
and thus lower compared to a sterile neutrino in thermal equilibrium with $\langle p \rangle_T =3.15 T_N$. We find that the sterile neutrinos become non-relativistic at a temperature
\begin{equation}
T_{N,nr}\simeq \frac{m_N}{2.46}\;.
\end{equation}
The corresponding temperature of the SM thermal bath $T_{nr}$ is obtained using the usual entropy dilution given in Eq.~\eqref{eq:TN2T} with $g_*^s(T_d)=g_*^\rho(T_d)=110.75$ and $g_*^s(T_{nr})=3.94$ and thus the time $t_{nr}$, when the sterile neutrinos become non-relativistic, is determined by 
\begin{equation}
t_{nr}=\frac{M_0}{2T_{nr}^2}
\simeq \frac{1.82}{\sqrt{g_*^{\rho}(T_d)}}\left(\frac{g_*^s(T_{nr})}{g_*^s(T_d)}\right)^{2/3}\frac{M_{Pl}}{ m_N^2}\simeq 1500 s \left(\frac{10\, \mathrm{keV}}{m_N}\right)^2\;. \label{entropytnr} 
\end{equation}
Thus we can finally evaluate the integral for the free-streaming horizon by evaluating%
\begin{figure}[tb]\centering
\includegraphics[width=0.7\linewidth]{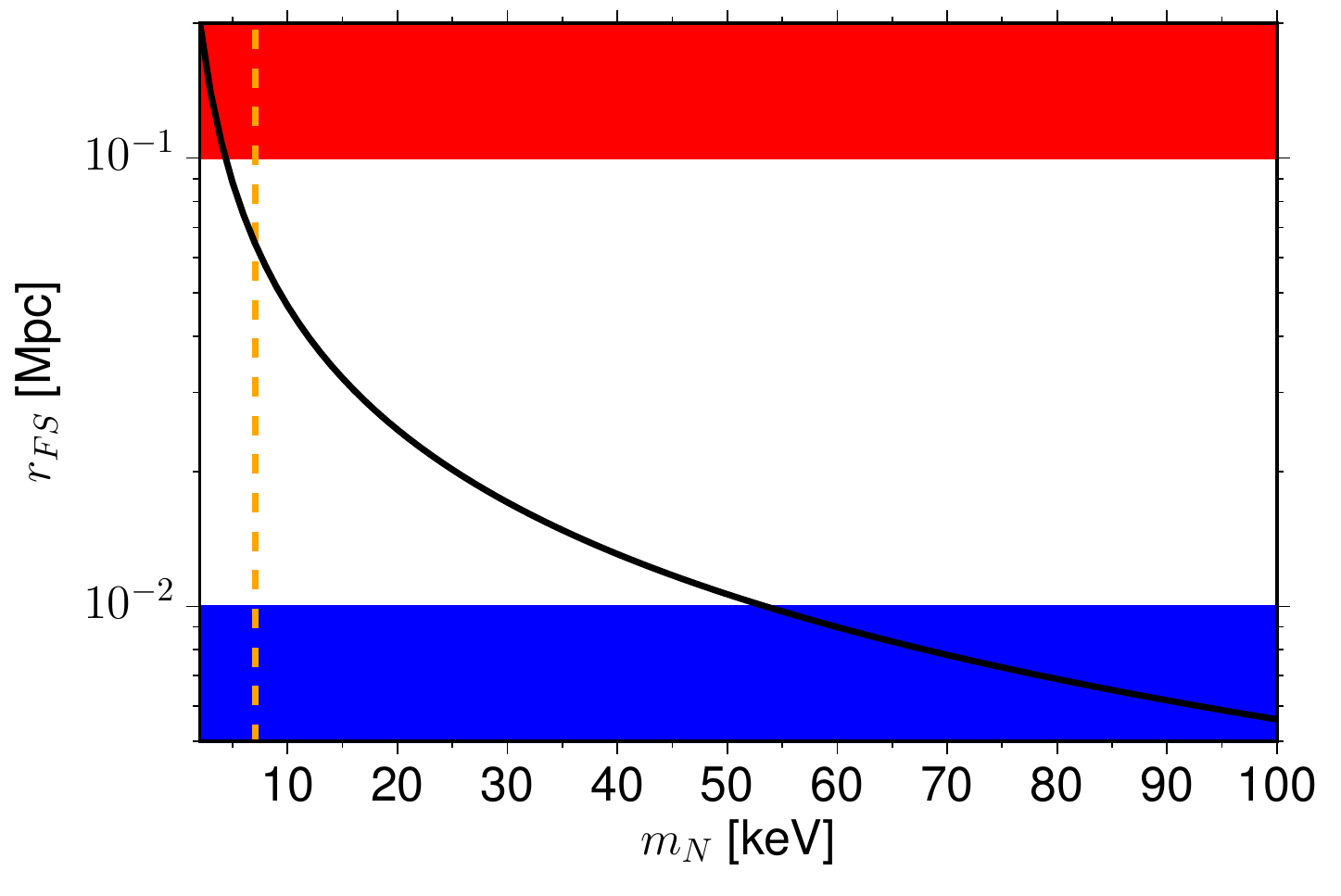}
\caption{Free-streaming horizon vs.~keV sterile neutrino mass. The regions of hot and cold dark matter are marked red and blue, respectively. A sterile neutrino mass $m_N=7.1$ keV is indicated by the dashed orange line.}
\label{FreeStream}
\end{figure}
it piece-wise in the different regions set by $t_{nr}$ and $t_{eq}$, respectively
\begin{equation}
r_{\rm FS}= \frac{\sqrt{t_{eq} t_{nr}}}{a_{eq}} \left(5+\ln\frac{t_{eq}}{t_{nr}}\right)
\simeq 0.047\,\mathrm{Mpc} \left(\frac{10\,\mathrm{keV}}{m_N}\right)\;.
\end{equation}
where $t_{eq}=1.9 \times 10^{11} \mathrm{s}$ and $a_{eq}=8.3 \times 10^{-5}$. A detailed derivation can be found in Refs.~\cite{Hasenkamp:2012ii,Merle:2013wta,Adulpravitchai:2014xna}. Note that we do not have the additional entropy dilution factor, because it is already included in $t_{nr}$ in \Eqref{entropytnr}. Note, that the free-streaming horizon does not (strongly) depend on the mass of the heavy scalar or the effective coupling, which determine the sterile neutrino abundance. There is only an implicit dependence via the effective degrees of freedom at decay. A different scalar mass will lead to a different decay temperature, but for scalar masses above $100$ GeV, the number of effective degrees of freedom stays almost constant.

We show the free-streaming horizon in Fig.~\ref{FreeStream}. The red region indicates when the free-streaming horizon becomes larger than $0.1$ Mpc and the keV sterile neutrinos can be considered as hot dark matter. The blue region indicates the region when keV sterile neutrinos are in the cold dark matter regime. We take as benchmark value $r_{FS}\lesssim 0.01$ Mpc following Refs.~\cite{Merle:2013wta,Adulpravitchai:2014xna}.

We find free-streaming horizons within the desirable range for warm dark matter for a sizeable fraction of the parameter space. More precisely, the free-streaming horizon is in the warm dark matter range for keV sterile neutrino masses of $4$ to about $53$ keV. For example for  $m_N=7.1$ keV, the keV sterile neutrino mass fitting the recently claimed X-ray line observation at 3.55 keV~\cite{Bulbul:2014sua,Boyarsky:2014jta}, we obtain $r_{FS} \simeq 0.06$ Mpc, which is well within the range of warm dark matter.

\section{Effective Degrees of Freedom}\label{sec:Neff}
With the introduction of an additional sterile neutrino, one might wonder about its contribution to the effective relativistic degrees of freedom. Its contribution can be quantified by\cite{Merle:2015oja}
\begin{equation}
\Delta N_\mathrm{eff}= \frac{\rho_N-m_N n}{\rho_{1\nu}}\qquad\mathrm{with}\quad\rho_{1\nu}=\frac74 \frac{\pi^2}{30}T_\nu^4\;,
\end{equation}
where we subtracted the non-relativistic energy density contained in the mass of the sterile neutrino. The energy density $\rho_N$ is given in Eq.~\eqref{eq:rhoN}. Thus we find that there are no additional relativistic degrees of freedom in the non-relativistic limit $r_N\equiv m_N/T_N\gg \hat x$, but in the ultra-relativistic limit ($r_N\ll \hat x$) the additional effective relativistic degrees of freedom are
\begin{equation}
\Delta N_\mathrm{eff}(T)
=0.0379\frac{ M_{Pl}}{m_{kk} \sqrt{g_*^\rho(T_{d})}} \left(1-\frac{r_N}{2.46}\right)  \left(\frac{g_*^s(T)}{g_*^s(T_{d})}\right)^{4/3}\left(\frac{T}{T_\nu}\right)^{4} \sum_\alpha |y_{LN,\alpha}|^2\;.
\end{equation}
Hence we find for a temperature $T_\nu=T_\mathrm{BBN}\simeq 4$ MeV shortly before the onset of big bang nucleosynthesis $\Delta N_\mathrm{eff}(T_\mathrm{BBN})\sim 10^{-3}$ with $g_*^s(T_\mathrm{BBN})=10.75$, $g_*^s(T_{d})=110.75$,  $y_{LN,\alpha}\sim 10^{-8}$, the scalar mass $m_{kk}\sim500$ GeV, $m_N\sim 10$ keV and hence a negligibly small contribution to the effective number of neutrinos. At recombination the temperature is well below the mass of the keV sterile neutrinos, $m_N$, i.e. $r_N\gg \hat x$, and thus $\Delta N_\mathrm{eff}(T_\mathrm{rec})\simeq0$.

\section{Sterile Neutrino Mixing and the X-ray Line}\label{sec:higgsed}
The scalar doublet scalar $H_\nu$ may obtain a vev, $\braket{H_\nu}=\begin{pmatrix}0& v_\nu\end{pmatrix}^T$, similar to the SM Higgs. 
A vev will induce mixing between the sterile neutrino and the active neutrinos
\begin{equation}
\theta_\alpha \simeq  \frac{y_{LN,\alpha} v_\nu}{m_N}\;.
\end{equation}
and lead to a new contribution to the active neutrino mass via the seesaw mechanism 
\begin{equation}\label{eq:ContrNu}
\Delta m_{\nu,\alpha\beta} = m_N \sin\theta_\alpha \sin\theta_\beta\;.
\end{equation} 
The mixing allows the production of the sterile neutrinos via neutrino oscillations~\cite{Barbieri:1990vx,Enqvist:1990ad,Dodelson:1993je}, which can be estimated using the approximate formula~\cite{Kusenko:2009up}
\begin{equation}
\Omega_{N,osc}h^2\simeq 0.2\times \frac{\sum_\alpha\sin^2\theta_\alpha}{3\times 10^{-9}}\left(\frac{m_N}{3 \mathrm{keV}}\right)^{1.8}\;.
\end{equation}
In addition the sterile neutrino can decay into a photon and an active neutrino generating an X-ray line. This already constrains the mixing angle. See Ref.~\cite{Horiuchi:2013noa} for different constraints on a keV sterile neutrino which is produced via oscillations. 
Last year two independent groups observed an X-ray line at 3.55 keV~\cite{Bulbul:2014sua,Boyarsky:2014jta}, which can be explained by the decay of sterile neutrino DM to a photon and a neutrino with an active-sterile mixing, $\sum_{\alpha} \sin^2(2\theta_\alpha) \simeq 7 \times 10^{-11}$. However, the observation is still debated~\cite{Boyarsky:2014ska,Boyarsky:2014paa,Riemer-Sorensen:2014yda,Jeltema:2014qfa,Malyshev:2014xqa}.

The vev $v_\nu$ can be naturally small and satisfy the X-ray bound, if it is induced via a small, possibly complex, $Z_2$ soft-breaking term~\cite{Ma:2000cc,Haba:2014taa}, 
\begin{equation}
V_{\rm soft}= (\mu_{12}^2 H^\dagger H_\nu+\hc)
\end{equation}
after the electroweak symmetry is broken. We obtain for the vev $v_\nu$
\begin{align}
\mathrm{Re}(v_\nu) &\simeq -\frac{\mathrm{Re}\left(\mu_{12}^2\right)v}{m_2^2+v^2\left(\lambda_3+\lambda_4+\lambda_5\right)} & 
\mathrm{Im}(v_\nu) &\simeq \frac{\mathrm{Im}\left(\mu_{12}^2\right)v}{m_2^2+v^2\left(\lambda_3+\lambda_4-\lambda_5\right)}\;.
\end{align}
There are no charge-breaking minima at leading order in $\mu_{12}^2$. The scalar masses will receive small corrections proportional to $\mu_{12}^2$ appearing at the second order in the scalar mass in \Eqref{masses}. 
\begin{figure}[tb]\centering
\includegraphics[width=0.7\linewidth]{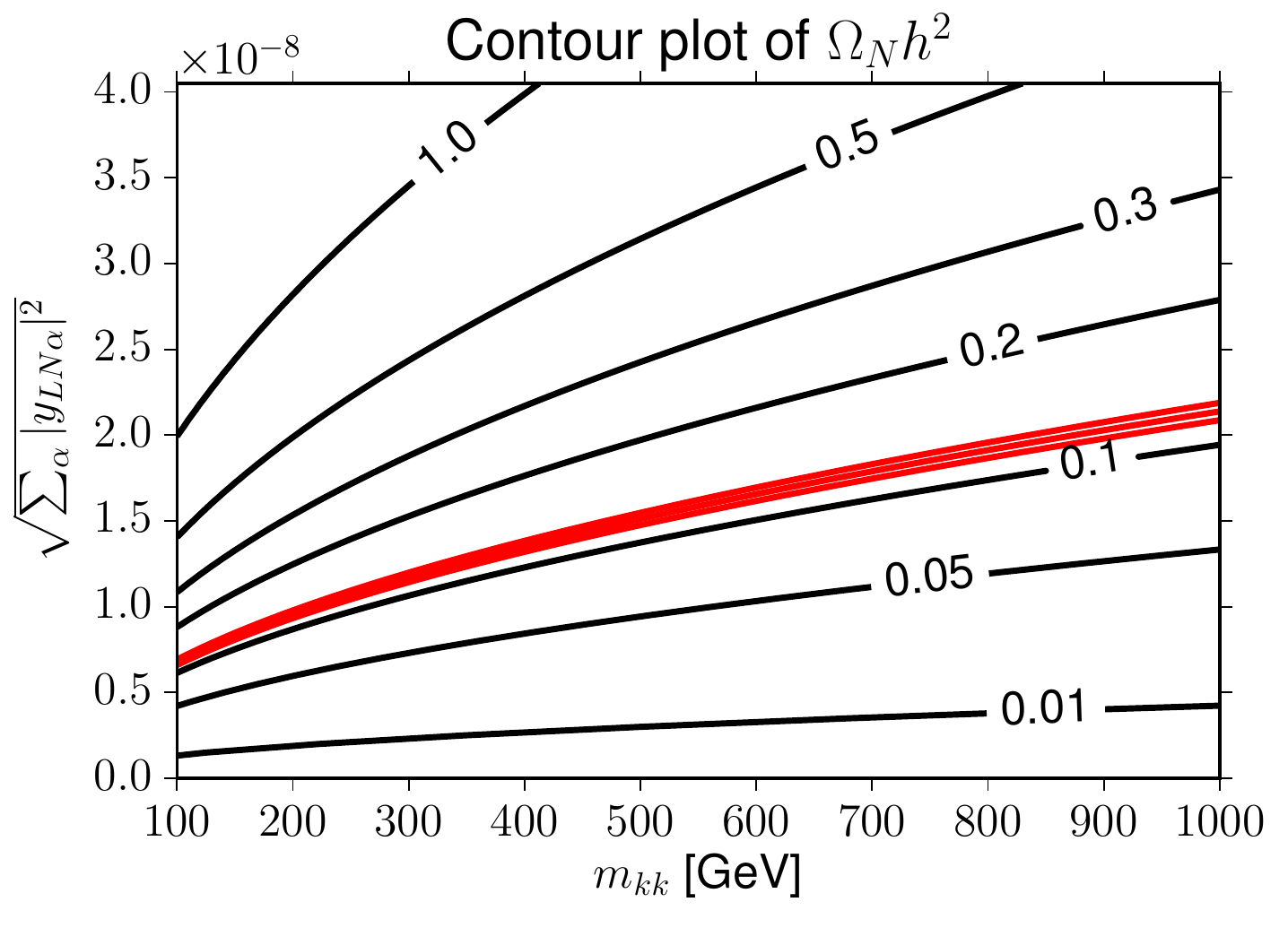}
\caption{Contour plot showing the dark matter abundance $\Omega h^2$ for the benchmark point with $m_N=7.1$ keV and $\sum_{\alpha} \sin^2(2\theta_\alpha) \simeq 7 \times 10^{-11}$. The red band indicates the $2\sigma$-allowed region around the measured by Planck~\cite{Ade:2015xua}.}
\label{fig:DM7keV}
\end{figure}

As the production via non-resonant neutrino oscillations can only give a subdominant contribution to the abundance of keV sterile neutrinos~\cite{Horiuchi:2013noa}\footnote{This does not apply to sterile neutrino production via resonant neutrino oscillations~\cite{Shi:1998km}. Thus the $3.55$ keV X-ray line can be explained in models like in Ref.~\cite{Canetti:2012kh,1506.06752v1}.}, the production is dominated by scalar decay. Thus we can translate the limit on the mixing angle from X-ray observations into a limit on the vev, $v_\nu$. We find $v_\nu\lesssim 1-10$ MeV depending on the sterile neutrino mass $m_N\simeq 2-100$ keV and the scalar mass $m_{kk}\simeq 100-1000$ GeV. Hence the contribution to the active neutrino masses in Eq.~\eqref{eq:ContrNu} is much smaller than the solar mass scale. The small splitting might explain the existence of pseudo-Dirac neutrinos, if the active neutrino mass originates from a Dirac mass term. 
Taking the claimed hint for an X-ray line at 3.55 keV~\cite{Bulbul:2014sua,Boyarsky:2014jta}, we show in Fig.~\ref{fig:DM7keV} the dark matter abundance as a function of the scalar mass $m_{kk}$ and the effective Yukawa coupling $\sqrt{\sum_\alpha |y_{LN,\alpha}|^2}$ fixing $m_N=7.1$ keV and the active-sterile mixing $\sum_{\alpha} \sin^2(2\theta_\alpha) \simeq 7 \times 10^{-11}$. The red band indicates the $2\sigma$-allowed region around the best-fit value measured by Planck~\cite{Ade:2015xua}. The required effective Yukawa couplings are of order $(5 - 22) \times 10^{-9}$ for scalar masses $m_{kk}\simeq 100-1000$ GeV.

\section{Conclusions}\label{sec:conclusions}

In the Standard Model of particle physics extended by one singlet Majorana fermion, keV sterile neutrinos are usually produced via neutrino oscillations and the decay of the electroweak Higgs doublet scalar into the keV sterile neutrino is a subdominant contribution. However, we showed in this paper, that the decay of an electroweak doublet scalar can become the dominant production mechanism in a two Higgs doublet model, if the vev of the electroweak scalar doublet is small or even vanishes. Neutrino oscillations will only account for a minor additional contribution to the keV sterile neutrino abundance. Thus we do not consider a vev of the electroweak scalar doublet for the first part of the paper and only indicate in Sec.~\ref{sec:higgsed} the possible changes, when the electroweak doublet scalar obtains a vev. As long as the vev is small enough, the produced abundance can be approximated by the result for a vanishing vev. The vev will lead to a mixing of the sterile neutrino with the active neutrinos. This renders it unstable via a two body decay allowing to search for the keV sterile neutrinos using X-ray line searches.

We explicitly derived an analytic expression for the momentum distribution of the keV sterile neutrino for late times, studied its free-streaming horizon and briefly commented on its contribution to $N_\mathrm{eff}$, the effective number of neutrinos in the early Universe, which is neglibly small.
This production mechanism leads to a cooler spectrum of the sterile neutrino. The range for the sterile neutrino to be the warm dark matter is in between 4 and 53 keV.

The mechanism to produce sterile  neutrinos via the decay of a Higgs doublet can be easily embedded in models of neutrino mass, which naturally explain the smallness of neutrino mass. This requires the addition of two or more sterile neutrinos. If the second Higgs doublet obtains a tiny vev, neutrino mass is naturally suppressed, while the Yukawa couplings of the other neutrinos can be relatively sizeable~\cite{Ma:2000cc,Haba:2014taa}. For a vanishing vev of the second Higgs doublet, neutrino masses can be induced via the radiative seesaw in the scotogenic model~\cite{Ma:2006km} and dark matter is produced via freeze-in~\cite{Molinaro:2014lfa}.

\vspace{2ex}
{\em Note added:} A recent publication~\cite{Drewes:2015eoa} pointed out additional
thermal corrections to the production rate, which become relevant at high
temperatures. We do not expect any significant corrections, because keV sterile
neutrinos are dominantly produced at low temperatures, $T\lesssim m_X$. The
inclusion of these effects goes beyond the scope of this work,
which focused on a simple analytic discussion of this novel keV sterile neutrino
production mechanism. A future numerical discussion should include these effects
and also include the other neglected subdominant corrections discussed in Sec.
3. 

\section*{Acknowledgements}
We would like to thank Andreas Hohenegger for very helpful discussions about Boltzmann equations. AA would like to thank Piyabut Burikham for his helpful comments. This work was supported by the Australian Research Council and Chulalongkorn University through Ratchadapisek Sompote Endowment Fund.
 
\appendix

\section{Decay Widths}\label{app:decaywidth}
The decay width of the scalar fields are given as 
\begin{eqnarray}
\Gamma(k\to N_1 \nu_\alpha) &=& \frac{m_k |y_{LN,\alpha}|^2}{32 \pi}(1-\frac{m_N^2}{m_k^2})^2 \simeq  \frac{m_k |y_{LN,\alpha}|^2}{32 \pi} \\
\Gamma(K^0\to N_1 \nu_\alpha) &=& \frac{m_{K^0} |y_{LN,\alpha}|^2}{32 \pi}(1-\frac{m_N^2}{m_{K^0}^2})^2 \simeq  \frac{m_{K^0} |y_{LN,\alpha}|^2}{32 \pi} \\
\Gamma(K^+\to N_1 \bar{l_\alpha}) &=& \frac{m_{K^+} |y_{LN,\alpha}|^2}{16 \pi}(1-\frac{m_N^2}{m_{K^+}^2})^2 \simeq  \frac{m_{K^+} |y_{LN,\alpha}|^2}{16 \pi} \;.
\end{eqnarray}
Hence the decay widths in the limit $m_N\ll m_X$ can be approximated by 
\begin{equation}\label{eq:GammaM}
\frac{\Gamma(k\to N_1 \nu_\alpha)}{g_k m_k} \simeq  
\frac{\Gamma(K^0\to N_1 \nu_\alpha)}{g_{K^0} m_{K^0}} \simeq
\frac{\Gamma(K^+\to N_1 \bar{l_\alpha})}{ g_{K^+}m_{K^+}}\simeq
\frac{ |y_{LN,\alpha}|^2}{32 \pi}\;.
\end{equation}

\section{Dark Matter Momentum Distribution Function}\label{app:distrib}

In case of the production of sterile neutrinos from decay of the real scalar $k$, the collision term is given by
\begin{align}
C[f_N]\equiv \frac{1}{2E_N}\int d\Pi_k d\Pi_\nu (2\pi)^4\delta^{(4)}(p_k-p_N-p_\nu) \overline{| \mathcal{M}|}^2\left(f_k (1-f_N)(1-f_\nu)-f_N f_\nu (1+f_k)\right)\;,
\end{align}
with the Lorentz-invariant phase space element
\begin{equation}
d\Pi=\frac{g d^3 p}{(2\pi)^3 2E}\;,
\end{equation}
which is denoted $d\Pi_\nu$ ($d\Pi_k$) for neutrinos (the real scalar). The number of degrees of freedom are $g_k=1$ for the scalar and $g_\nu=2$ for the neutrino. The distribution functions of the scalar (neutrino) are $f_k$ ($f_\nu$) and the spin-averaged matrix element is given by
\begin{equation}
\overline{|\mathcal{M}|}^2= \frac14 |y_{LN}|^2 p_\nu \cdot p_N = \frac14 |y_{LN}|^2 \left(m_k^2-m_N^2-m_\nu^2\right)\;.
\end{equation}
Using the decay width $\Gamma_k \simeq |y_{LN}|^2 m_k/32\pi$ in the limit of negligible final state masses, $m_N=m_\nu=0$, we can rewrite the matrix element 
\begin{equation}
\overline{|\mathcal{M}|}^2= \frac{8\pi \Gamma_k}{m_k} m_k^2
\end{equation}
and generalise it to any of the three considered scalars using Eq.~\eqref{eq:GammaM}
\begin{equation}
\overline{|\mathcal{M}|}^2= 8\pi g_X m_X\Gamma_X\;.
\end{equation}
Neglecting all terms proportional to $f_N$ and taking the ultra-relativistic approximation $E_N\simeq p_N$, we obtain
\begin{equation}
C[f_N]\equiv \frac{4\pi g_Xm_X\Gamma_X}{p_N} \int d\Pi_X d\Pi_\nu (2\pi)^4\delta^{(4)}(p_X-p_N-p_\nu) f_X \left(1-f_\nu\right) \;.
\end{equation}
Note that we do not neglect Pauli-blocking, i.e. do not approximate $1-f_\nu\simeq 1$.
The integration over the neutrino momenta $p_\nu$ is easily evaluated using the $\delta$-function of the momenta. 
As the integrand only depends on $E_\nu$ and thus the scalar product $\vec p_X \cdot \vec p_N$, which can be evaluated via the delta function of the energies
\begin{equation}
E_X-E_N=E_\nu=|\vec p_\nu|=\sqrt{\vec p_X^2 +\vec p_N^2-2\vec p_X \cdot \vec p_N}\;,
\end{equation}
the collision term can we written as
\begin{equation}
C[f_N]=\frac{g_X m_X \Gamma_X}{2p_N^2}  \int_{p_N+\frac{m_X^2}{4p_N}}^\infty d E_X f_X\left(\frac{E_X}{T}\right) \left(1-f_\nu\left(\frac{E_X}{2T}\right)\right)\;.
\end{equation}
Using the dimensionless variables 
\begin{align}
x_{N,X}&=\frac{p_{N,X}}{T},&
r&=\frac{m_X}{T},&
r_N&=\frac{m_N}{T}
\end{align}
we rewrite the Liouville operator in the radiation dominated epoch assuming $g_*^\rho$ to be constant
\begin{equation}
L[f_N]=  H r \frac{\partial f_N}{\partial r} 
\end{equation}
and the collision term 
\begin{equation}
C[f_N]=g_X \Gamma_X \frac{r}{2x_N^2} \int_{x_N+\frac{r^2}{4x_N}}^\infty f_X(y_X)\left(1-f_\nu\left(\frac{y_X}{2}\right)\right) d y_X\;,
\end{equation}
where we use the Friedmann equation in a radiation dominated epoch to express the Hubble constant as $H(T)=T^2 / M_0 = m_X^2/M_0 r^2$ where $M_0$ is defined in Eq.~\eqref{eq:M0}.
Using our result from the collision term of a scalar decay, we find the Boltzmann equation for the distribution function $f_N$ of the sterile neutrinos
\begin{equation}
\frac{\partial f_N}{\partial r} =  \frac{M_0 g_X \Gamma_X}{2m_X^2}
 \frac{ r^2}{x_N^2} \int_{x_N+\frac{r^2}{4x_N}}^\infty d y_X f_X(y_X)\left(1-f_\nu\left(\frac{y_X}{2}\right)\right)\;.
\end{equation}
Hence the distribution function can be obtained by a simple integration
\begin{equation}\label{eq:fNbeforeInt}
f_N(x_N,r) = \frac{M_0 g_X\Gamma_X}{2m_X^2}
  \int_0^r d r^\prime \frac{r^{\prime2} }{x_N^2} \int_{x_N+\frac{r^{\prime2}}{4x_N}}^\infty d y_X  f_X(y_X) \left(1-f_\nu\left(\frac{y_X}{2}\right)\right)\;.
\end{equation}
If the scalar and the neutrino are in thermal equilibrium, they are described by a Bose-Einstein distribution function $f_{BE}^{-1}(y) =e^{y}-1$ and Fermi-Dirac distribution $f_{FD}^{-1}(y)=e^y+1$, respectively. We can perform the integral over $y_X$,
\begin{equation}
f_N(x_N,r) = \frac{M_{0}g_X\Gamma_X}{2m_X^2}
  \int_0^r d r^\prime \frac{r^{\prime2} }{x_N^2} g\left(e^{\frac12\left(x_N+\frac{r^{\prime2}}{4x_N}\right)}\right)
\end{equation}
with 
\begin{equation}\label{eq:g}
g(z)=-\frac{1}{1+z}-\frac12\ln\left(\frac{z-1}{z+1}\right)\;.
\end{equation}
Changing the variables of integration to $y=r^{\prime2}/8x_N$ we obtain
\begin{equation}\label{eq:fNint}
f_N(x_N,r) = \frac{4\sqrt{2}M_{0}g_X\Gamma_X}{m_X^2} \frac{1}{\sqrt{x_N}}
  \int_0^{r^2/8x_N} y^{\frac12} g\left(e^{\frac{x_N}{2}} e^y\right) dy
\end{equation}
We will employ the integral representation of the incomplete polylogarithm\footnote{The integral representation of the incomplete polylogarithm is also known as the incomplete Bose-Einstein and Fermi-Dirac integrals, more precisely
\begin{equation}
F_j(x,b)\equiv \frac{1}{\Gamma(j+1)}\int_b^\infty \frac{t^j}{e^{t-x}+1} dx =-\mathrm{Li}_{j+1}\left(b,-e^x\right)\;.
\end{equation}}
\begin{align}\label{eq:polylog}
\mathrm{Li}_s(b,z)&=\frac{1}{\Gamma(s)}\int_b^\infty \frac{t^{s-1}}{e^t/z-1}dt && \textrm{all z but Re(z)}\geq 1\\\nonumber
\mathrm{Li}_s(b,-z)&=-\frac{1}{\Gamma(s)}\int_b^\infty \frac{t^{s-1}}{e^t/z+1}dt && \textrm{all z but Re(z)}\leq -1\;,
\end{align}
which are well-defined for Re$(s)>0$,
to rewrite the integral in terms of known functions. The usual polylogarithm is obtained for $b=0$, i.e. $\mathrm{Li}_s(z)\equiv\mathrm{Li}_s(0,z)$. 
The first summand of Eq.~\eqref{eq:g} can be integrated directly using the integral representation of the polylogarithm in Eq.~\eqref{eq:polylog}, while the second term requires integration by parts to obtain fractions, which can then be integrated using Eq.~\eqref{eq:polylog}. Finally the distribution function can be expressed at late times with $r^2\gg 8 x_N$ by
\begin{align}
f_N^0(x_N) & = \frac{\sqrt{2\pi}M_{0}g_X\Gamma_X}{m_X^2} \frac{1}{\sqrt{x_N}} \left[
2 \mathrm{Li}_{\frac32}\left(-e^{-\frac{x_N}{2}}\right)
+\mathrm{Li}_{\frac52}\left(e^{-\frac{x_N}{2}}\right)
-\mathrm{Li}_{\frac52}\left(-e^{-\frac{x_N}{2}}\right)
\right]
\end{align}
A comparison to the result without Pauli-blocking, 
\begin{equation}
f_N^{0,nP}(x_N)=\frac{\sqrt{\pi}M_{0}g_X\Gamma_X}{2 m_X^2} \frac{1}{\sqrt{x_N}} \mathrm{Li}_{\frac52}\left(e^{-x_N}\right)
\end{equation}
shows that the Pauli-blocking leads to the correction
\begin{align}
f_N^0(x_N) -f_N^{0,nP}(x_N)& = \frac{\sqrt{2\pi}M_{0}g_X\Gamma_X}{2 m_X^2} \frac{1}{\sqrt{x_N}} \left[
2 \mathrm{Li}_{\frac32}\left(-e^{-\frac{x_N}{2}}\right)
-\mathrm{Li}_{\frac52}\left(e^{-\frac{x_N}{2}}\right)
-3\mathrm{Li}_{\frac52}\left(-e^{-\frac{x_N}{2}}\right)
\right]\;.
\end{align}
As expected the correction is negative and becomes larger for smaller $x_N$.  This agrees with the distribution function given in Ref.~\cite{Boyanovsky:2008nc}.
In the limit $x_N\gg 0$ we obtain the Maxwell-Boltzmann result
\begin{equation}
f_N^0(x_N)  \stackrel{x_N\gg0}{\simeq} \frac{\sqrt{\pi}M_{0}g_X\Gamma_X}{m_X^2} \frac{e^{-x_N}}{\sqrt{x_N}}\;,
\end{equation}
which agrees with the result obtained from taking the Maxwell-Boltzmann limit in Eq.~\eqref{eq:fNbeforeInt}.
\begin{equation}
f_N^{MB}(x_N,r ) = \frac{\sqrt{\pi} M_0 g_X\Gamma_X}{ m_X^2} \frac{e^{-x_N}}{\sqrt{x_N}} \left(1-\frac{\Gamma\left(\frac32,\frac{r^2}{4x_N}\right)}{\Gamma\left(\frac32\right)}\right)   \;,
\end{equation}
where we used the incomplete $\Gamma$ function 
\begin{equation}
\Gamma(s,x) \equiv \int_x^\infty t^{s-1} e^{-t} dt\;.
\end{equation}
The final expression for the distribution function at a finite temperature is given by
\begin{equation}
f_N(x_N,r) = f_N^0(x_N) - \frac{\sqrt{2\pi}M_{0}g_X\Gamma_X}{m_X^2} \frac{1}{\sqrt{x_N}} h\left(\frac{r^2}{8x_N},e^{-\frac{x_N}{2}}\right)
\end{equation}
with the function
\begin{equation}
h(b,z)= 2 \mathrm{Li}_{\frac32}\left(b,z\right)
+\mathrm{Li}_{\frac52}\left(b,z\right)
-\mathrm{Li}_{\frac52}\left(b,-z\right)
- \frac{4}{3\sqrt{\pi}}b^{\frac32}\ln\left(\frac{e^b-z}{e^b+z}\right)\stackrel{b\to\infty}{\longrightarrow}0\;.
\end{equation}
We are mainly interested in the distribution function after the sterile neutrino DM production has finished and thus we will use the zero-temperature limit $f_N^0(x_N)$ in the following.

The generalisation to multiple scalar fields is simply the sum of the different contributions 
\begin{align}\label{eq:fNlate}
f_N^0(x_N) & = \frac{\sqrt{90}}{2\pi}  \sum_X\frac{g_X\Gamma_XM_{pl}}{m_X^2 \sqrt{g_*^\rho(T_{d,X})}}
 \frac{1}{\sqrt{x_N}} \left[ 
2 \mathrm{Li}_{\frac32}\left(-e^{-\frac{x_N}{2}}\right)
+\mathrm{Li}_{\frac52}\left(e^{-\frac{x_N}{2}}\right)
-\mathrm{Li}_{\frac52}\left(-e^{-\frac{x_N}{2}}\right)
\right]
\;.
\end{align}
The function $x^2 f_N^0(x)$ has a maximum at $\hat x\simeq 1.54$ and falls off very quickly away from the maximum. A typical value for the momentum is thus of order $\hat x$.
After the sterile neutrinos are produced via the decay, they are completely decoupled from the thermal bath of the SM particles and generally have a different temperature $T_N$ compared to the thermal bath of SM particles. Obviously the distribution function of the sterile neutrinos does not resemble a thermal Maxwell-Boltzmann, Fermi-Dirac or Bose-Einstein statistic, but represents a non-equilibrium distribution function.

\section{Temperature of the Sterile Neutrino Dark Matter}\label{app:TempSterileNu}
 In the following we will calculate all relevant quantities for the sterile neutrino dark matter sector for sufficiently late time and low temperature $T_N\ll m_X/2x_N$ using the distribution function in Eq.~\eqref{eq:fNlate}. In order to simplify the notation we will also drop the subscript $N$ from $p_N$ and $x_N=p_N/T_N$. 
The number density and the average momentum are given by
\begin{align}\label{eq:numDensity}
n & = g_N\int \frac{d^3p}{(2\pi)^3} f_N(p) =  \frac{T_N^3}{2\pi^2} \int_0^\infty x^2 f_N(x) dx\\\nonumber
&=
\frac{9\sqrt{5}}{2\pi^{5/2}}\left(\zeta(5)-\frac{13}{16}\zeta(4)\right) \sum_X \frac{g_X\Gamma_XM_{pl}}{m_X^2\sqrt{ g_*^{\rho}(T_{d,X})}} T_N^3
\simeq 0.091 \sum_X \frac{g_X\Gamma_XM_{pl}}{m_X^2\sqrt{ g_*^{\rho}(T_{d,X})}}  T_N^3 \\
 \braket{p} & = \frac{g_N}{n}\int \frac{d^3p}{(2\pi)^3}p f_N(p) 
=  \frac{T_N^4}{2n\pi^2 } \int_0^\infty x^3 f_N(x) dx
= 5 \frac{29\zeta(5)-32\zeta(6)}{26\zeta(4)-32\zeta(5)}T_N \simeq 2.46 T_N\;,
\end{align}
where we used the integral formulas
\begin{align}
\int_0^\infty x^n \mathrm{Li}_s(e^{-x/a})dx &=a^{n+1}\Gamma(n+1) \zeta(n+s+1)\\
\int_0^\infty x^n \mathrm{Li}_s(-e^{-x/a})dx &=a^{n+1} \left(-1+2^{-n-s}\right)\Gamma(n+1) \zeta(n+s+1)
\end{align}
to evaluate the integrals. The result for the number density in Eq.~\eqref{eq:numDensity} is consistent with the calculation of the yield $Y=n/s$ in Eq.~\eqref{eq:Yinf1} using the integrated Boltzmann equation in the Maxwell-Boltzmann approximation.
Using the expression for the number density and the distribution function we can determine how the temperature of the sterile neutrinos depends on the scale factor~\cite{Weinberg:2008xx}. The sterile neutrinos are not interacting and thus the only effect is due to the cosmic expansion. 
After the abundance of the sterile neutrino is frozen-in, $n(x)a^3$ does not change and we can relate the number density $n$ at time $t$ to the number density $n^\prime$ at a later time $t^\prime$, $n(x)a^3=n'(x')(a^\prime)^3$. Then we obtain from the number density of the sterile neutrinos with momenta between $p$ and $p+dp$ the following relation between the distribution functions at different times
\begin{align}
T_N^3 x^2 f_N(x) dx  = \left(\frac{a^\prime}{a}\right)^3 T_N^{\prime 3} x^{\prime2} f_N^\prime (x^\prime) dx^\prime =   T_N^3 x^2 f_N^\prime ( \frac{a}{a^\prime} \frac{T_N}{T_N^\prime} x) dx
\end{align}
where we used the dependence of the momentum  on the scale factor, $p\propto a^{-1}$, to relate $x$ and 
\begin{equation}
x^\prime = \frac{p^\prime}{T^\prime_N}=\frac{a}{a^\prime} \frac{p}{T_N} \frac{T_N}{T_N^\prime} = \frac{a}{a^\prime} \frac{T_N}{T_N^\prime} x\;.
\end{equation}
Thus the distribution function $f^\prime_N$ at a later time $t^\prime$ has the same form with a temperature
\begin{equation}\label{eq:TaRel}
T_N^\prime = T_N \frac{a}{a^\prime}\;.
\end{equation}
The expression for the energy density is similarly obtained from 
\begin{equation}\label{eq:rhoN}
\rho_N = g_N\int \frac{d^3 p}{(2\pi)^3} E f_N(p) = \frac{T_N^4}{2\pi^2} \int_0^\infty x^2 \sqrt{x^2+r_N^2} f_N(x) dx =\begin{cases}
\braket{p} n & r_N\ll \hat x\\
m_N n & r_N \gg \hat x
\end{cases}
\end{equation}
with $r_N\equiv m_N/T_N$, where we took the ultra-relativistic or non-relativistic limit, respectively, in the last step.
The pressure is given by
\begin{align}
\mathcal{P}_N &= \left\langle \frac{p^2}{3 E(p)} \right\rangle 
= \frac{T_N^4}{6\pi^2} \int_0^\infty \frac{x^4}{\sqrt{x^2+r_N^2}} f_N(x) dx
=\begin{cases}
\frac13\braket{p} n  & r_N\ll \hat x\\
\frac76 \frac{61\zeta(6)-64\zeta(7)}{29\zeta(5)-32\zeta(6)}\frac{\rho_N}{r_N^{2}} 
& r_N \gg \hat x
\end{cases}
\end{align}
and thus the equation of state is
\begin{equation}
w=\frac{\mathcal{P}_N}{\rho_N} = 
\begin{cases}
\frac13 & r_N \ll \hat x\\
\frac76 \frac{61\zeta(6)-64\zeta(7)}{29\zeta(5)-32\zeta(6)}\frac1{r_N^{2}}\simeq \frac{1.16}{r_N^2}
  & r_N\gg \hat x
\end{cases}
\end{equation}
showing that the sterile neutrinos behave like matter for temperatures much below $m_N$, i.e. $r_N\gg 1$. Similarly the entropy density is given by
\begin{equation}
s_N=\frac{\rho_N+\mathcal{P}_N}{T_N}=\left(1+w\right)\frac{\rho_N}{T_N}=
\begin{cases}
\frac{20}{3}\frac{29\zeta(5)-32\zeta(6)}{26\zeta(4)-32\zeta(5)}n\simeq 3.28 n & r_N\ll \hat x\\
\left(1+w\right)r_N n  & r_N \gg \hat x
\end{cases}\;.
\end{equation}
Thus we find that the entropy density scales like $T_N^3$ in the ultra-relativistic limit for $r_N\ll \hat x$ and observe that the entropy of the sterile neutrino dark matter sector is separately conserved using Eq.~\eqref{eq:TaRel}.

We can use the conservation of entropy to obtain a relation between the temperature of the SM thermal bath $T$ and the temperature $T_N$ of the sterile neutrino dark matter sector.
Consequently we parameterize the entropy density similarly to the SM thermal bath
\begin{equation}
s=\frac{2\pi^2}{45} g_*^s T^3\quad\mathrm{and}\quad s_N=\frac{2\pi^2}{45} g_{*,N}^s T_N^3\;,
\end{equation}
we find for the entropy degrees of freedom in the sterile neutrino sector, $g_{*,N}^s$,
\begin{equation}
g_{*,N}^s=\sum_X\frac{5\pi^{3/2}}{56} \sqrt{\frac{5}{ g_*^{\rho}(T_{d,X})}}\frac{g_X\Gamma_X M_{Pl}}{ m_X^2}\;.
\end{equation}
Using  the conservation of entropy
\begin{equation}
g_*^s(T_d) T_d^3 a_d^3 = \left(g_*^s(T) T^3 + g_{*,N}^s T_N^3\right) a^3
\end{equation}
and assuming that all scalars decay at the same time $T_d$, we find for the temperature $T_N$,
\begin{equation}\label{eq:TN2T}
\frac{T_N}{T}=\left(\frac{g_{*}^s(T)}{g_{*}^s(T_d)-g_{*,N}^s}\right)^{1/3}
\simeq\left(\frac{g_{*}^s(T)}{g_{*}^s(T_d)}\right)^{1/3}\;.
\end{equation}
In the non-relativistic limit, we can employ Eq.~\eqref{eq:TaRel} to relate the temperature $T_N$ to the temperature $T_{N,nr}$,
\begin{equation}
T_N=T_{N,nr} \frac{a_{nr}}{a}
\end{equation}
and ultimately the temperature of the SM thermal bath $T$.

\bibliography{keV}

\providecommand{\href}[2]{#2}\begingroup\raggedright\begin{thebibliography}{10}

\bibitem{Abazajian:2012ys}
K.~Abazajian, M.~Acero, S.~Agarwalla, A.~Aguilar-Arevalo, C.~Albright, et~al.,
  {\it {Light Sterile Neutrinos: A White Paper}},
  \href{http://arxiv.org/abs/1204.5379}{{\tt arXiv:1204.5379}}.

\bibitem{Drewes:2013gca}
M.~Drewes, {\it {The Phenomenology of Right Handed Neutrinos}},  {\em
  Int.J.Mod.Phys.} {\bf E22} (2013) 1330019,
  [\href{http://arxiv.org/abs/1303.6912}{{\tt arXiv:1303.6912}}].

\bibitem{Kauffmann:1993gv}
G.~Kauffmann, S.~D. White, and B.~Guiderdoni, {\it {The Formation and Evolution
  of Galaxies Within Merging Dark Matter Haloes}},  {\em
  Mon.Not.Roy.Astron.Soc.} {\bf 264} (1993) 201.

\bibitem{Klypin:1999uc}
A.~A. Klypin, A.~V. Kravtsov, O.~Valenzuela, and F.~Prada, {\it {Where are the
  missing Galactic satellites?}},  {\em Astrophys.J.} {\bf 522} (1999) 82--92,
  [\href{http://arxiv.org/abs/astro-ph/9901240}{{\tt astro-ph/9901240}}].

\bibitem{Moore:2005jj}
B.~Moore, J.~Diemand, P.~Madau, M.~Zemp, and J.~Stadel, {\it {Globular
  clusters, satellite galaxies and stellar haloes from early dark matter
  peaks}},  {\em Mon.Not.Roy.Astron.Soc.} {\bf 368} (2006) 563--570,
  [\href{http://arxiv.org/abs/astro-ph/0510370}{{\tt astro-ph/0510370}}].

\bibitem{Kusenko:2006rh}
A.~Kusenko, {\it {Sterile neutrinos, dark matter, and the pulsar velocities in
  models with a Higgs singlet}},  {\em Phys.Rev.Lett.} {\bf 97} (2006) 241301,
  [\href{http://arxiv.org/abs/hep-ph/0609081}{{\tt hep-ph/0609081}}].

\bibitem{Kusenko:1997sp}
A.~Kusenko and G.~Segre, {\it {Neutral current induced neutrino oscillations in
  a supernova}},  {\em Phys.Lett.} {\bf B396} (1997) 197--200,
  [\href{http://arxiv.org/abs/hep-ph/9701311}{{\tt hep-ph/9701311}}].

\bibitem{Barbieri:1990vx}
R.~Barbieri and A.~Dolgov, {\it {Neutrino oscillations in the early universe}},
   {\em Nucl.Phys.} {\bf B349} (1991) 743--753.

\bibitem{Enqvist:1990ad}
K.~Enqvist, K.~Kainulainen, and J.~Maalampi, {\it {Refraction and Oscillations
  of Neutrinos in the Early Universe}},  {\em Nucl.Phys.} {\bf B349} (1991)
  754--790.

\bibitem{Dodelson:1993je}
S.~Dodelson and L.~M. Widrow, {\it {Sterile-neutrinos as dark matter}},  {\em
  Phys.Rev.Lett.} {\bf 72} (1994) 17--20,
  [\href{http://arxiv.org/abs/hep-ph/9303287}{{\tt hep-ph/9303287}}].

\bibitem{Horiuchi:2013noa}
S.~Horiuchi, P.~J. Humphrey, J.~Onorbe, K.~N. Abazajian, M.~Kaplinghat, et~al.,
  {\it {Sterile neutrino dark matter bounds from galaxies of the Local Group}},
   {\em Phys.Rev.} {\bf D89} (Nov., 2014) 025017,
  [\href{http://arxiv.org/abs/1311.0282}{{\tt arXiv:1311.0282}}].

\bibitem{Shi:1998km}
X.-D. Shi and G.~M. Fuller, {\it {A New dark matter candidate: Nonthermal
  sterile neutrinos}},  {\em Phys.Rev.Lett.} {\bf 82} (1999) 2832--2835,
  [\href{http://arxiv.org/abs/astro-ph/9810076}{{\tt astro-ph/9810076}}].

\bibitem{Shaposhnikov:2006xi}
M.~Shaposhnikov and I.~Tkachev, {\it {The nuMSM, inflation, and dark matter}},
  {\em Phys.Lett.} {\bf B639} (2006) 414--417,
  [\href{http://arxiv.org/abs/hep-ph/0604236}{{\tt hep-ph/0604236}}].

\bibitem{Bezrukov:2009yw}
F.~Bezrukov and D.~Gorbunov, {\it {Light inflaton Hunter's Guide}},  {\em JHEP}
  {\bf 1005} (2010) 010, [\href{http://arxiv.org/abs/0912.0390}{{\tt
  arXiv:0912.0390}}].

\bibitem{Petraki:2007gq}
K.~Petraki and A.~Kusenko, {\it {Dark-matter sterile neutrinos in models with a
  gauge singlet in the Higgs sector}},  {\em Phys.Rev.} {\bf D77} (2008)
  065014, [\href{http://arxiv.org/abs/0711.4646}{{\tt arXiv:0711.4646}}].

\bibitem{Frigerio:2014ifa}
M.~Frigerio and C.~E. Yaguna, {\it {Sterile Neutrino Dark Matter and Low Scale
  Leptogenesis from a Charged Scalar}},  {\em Eur.Phys.J.} {\bf C75} (2015) 31,
  [\href{http://arxiv.org/abs/1409.0659}{{\tt arXiv:1409.0659}}].

\bibitem{Merle:2015oja}
A.~Merle and M.~Totzauer, {\it {keV Sterile Neutrino Dark Matter from Singlet
  Scalar Decays: Basic Concepts and Subtle Features}},  {\em JCAP} {\bf 1506}
  (Feb, 2015) 011, [\href{http://arxiv.org/abs/1502.01011}{{\tt
  arXiv:1502.01011}}].

\bibitem{Hall:2009bx}
L.~J. Hall, K.~Jedamzik, J.~March-Russell, and S.~M. West, {\it {Freeze-In
  Production of FIMP Dark Matter}},  {\em JHEP} {\bf 03} (2010) 080,
  [\href{http://arxiv.org/abs/0911.1120}{{\tt arXiv:0911.1120}}].

\bibitem{Merle:2013wta}
A.~Merle, V.~Niro, and D.~Schmidt, {\it {New Production Mechanism for keV
  Sterile Neutrino Dark Matter by Decays of Frozen-In Scalars}},  {\em JCAP}
  {\bf 1403} (2014) 028, [\href{http://arxiv.org/abs/1306.3996}{{\tt
  arXiv:1306.3996}}].

\bibitem{Adulpravitchai:2014xna}
A.~Adulpravitchai and M.~A. Schmidt, {\it {A Fresh Look at keV Sterile Neutrino
  Dark Matter from Frozen-In Scalars}},  {\em JHEP} {\bf 1501} (Sep, 2015) 006,
  [\href{http://arxiv.org/abs/1409.4330}{{\tt arXiv:1409.4330}}].

\bibitem{Kang:2014cia}
Z.~Kang, {\it {FImP Miracle of Sterile Neutrino Dark Matter by Scale
  Invariance}},  \href{http://arxiv.org/abs/1411.2773}{{\tt arXiv:1411.2773}}.

\bibitem{Lello:2014yha}
L.~Lello and D.~Boyanovsky, {\it {Cosmological Implications of Light Sterile
  Neutrinos produced after the QCD Phase Transition}},  {\em Phys. Rev.} {\bf
  D91} (2015) 063502, [\href{http://arxiv.org/abs/1411.2690}{{\tt
  arXiv:1411.2690}}].

\bibitem{Abada:2014zra}
A.~Abada, G.~Arcadi, and M.~Lucente, {\it {Dark Matter in the minimal Inverse
  Seesaw mechanism}},  \href{http://arxiv.org/abs/1406.6556}{{\tt
  arXiv:1406.6556}}.

\bibitem{Shuve:2014doa}
B.~Shuve and I.~Yavin, {\it {Dark matter progenitor: Light vector boson decay
  into sterile neutrinos}},  {\em Phys. Rev.} {\bf D89} (2014), no.~11 113004,
  [\href{http://arxiv.org/abs/1403.2727}{{\tt arXiv:1403.2727}}].

\bibitem{Enqvist:2014zqa}
K.~Enqvist, S.~Nurmi, T.~Tenkanen, and K.~Tuominen, {\it {Standard Model with a
  real singlet scalar and inflation}},  {\em JCAP} {\bf 1408} (2014) 035,
  [\href{http://arxiv.org/abs/1407.0659}{{\tt arXiv:1407.0659}}].

\bibitem{Nurmi:2015ema}
S.~Nurmi, T.~Tenkanen, and K.~Tuominen, {\it Inflationary imprints on dark
  matter},  \href{http://arxiv.org/abs/1506.04048}{{\tt arXiv:1506.04048}}.

\bibitem{Bezrukov:2009th}
F.~Bezrukov, H.~Hettmansperger, and M.~Lindner, {\it {keV sterile neutrino Dark
  Matter in gauge extensions of the Standard Model}},  {\em Phys.Rev.} {\bf
  D81} (2010) 085032, [\href{http://arxiv.org/abs/0912.4415}{{\tt
  arXiv:0912.4415}}].

\bibitem{Nemevsek:2012cd}
M.~Nemevsek, G.~Senjanovic, and Y.~Zhang, {\it {Warm Dark Matter in Low Scale
  Left-Right Theory}},  {\em JCAP} {\bf 1207} (May, 2012) 006,
  [\href{http://arxiv.org/abs/1205.0844}{{\tt arXiv:1205.0844}}].

\bibitem{Bezrukov:2012as}
F.~Bezrukov, A.~Kartavtsev, and M.~Lindner, {\it {Leptogenesis in models with
  keV sterile neutrino dark matter}},  {\em J.Phys.} {\bf G40} (Apr., 2013)
  095202, [\href{http://arxiv.org/abs/1204.5477}{{\tt arXiv:1204.5477}}].

\bibitem{Tsuyuki:2014aia}
T.~Tsuyuki, {\it {Neutrino masses, leptogenesis, and sterile neutrino dark
  matter}},  {\em Phys.Rev.} {\bf D90} (2014) 013007,
  [\href{http://arxiv.org/abs/1403.5053}{{\tt arXiv:1403.5053}}].

\bibitem{Patwardhan:2015kga}
A.~V. Patwardhan, G.~M. Fuller, C.~T. Kishimoto, and A.~Kusenko, {\it {Diluted
  Equilibrium Sterile Neutrino Dark Matter}},
  \href{http://arxiv.org/abs/1507.01977}{{\tt arXiv:1507.01977}}.

\bibitem{Matsui:2015maa}
H.~Matsui and M.~Nojiri, {\it {Higgs sector extension of the neutrino minimal
  standard model with thermal freeze-in production mechanism}},
  \href{http://arxiv.org/abs/1503.01293}{{\tt arXiv:1503.01293}}.

\bibitem{Haba:2014taa}
N.~Haba, H.~Ishida, and R.~Takahashi, {\it {$\nu_R$ dark matter-philic Higgs
  for 3.5 keV X-ray signal}},  {\em Phys.Lett.} {\bf B743} (Jul, 2015) 35--38,
  [\href{http://arxiv.org/abs/1407.6827}{{\tt arXiv:1407.6827}}].

\bibitem{Molinaro:2014lfa}
E.~Molinaro, C.~E. Yaguna, and O.~Zapata, {\it {FIMP realization of the
  scotogenic model}},  {\em JCAP} {\bf 1407} (May, 2014) 015,
  [\href{http://arxiv.org/abs/1405.1259}{{\tt arXiv:1405.1259}}].

\bibitem{Ma:2006km}
E.~Ma, {\it {Verifiable radiative seesaw mechanism of neutrino mass and dark
  matter}},  {\em Phys. Rev.} {\bf D73} (2006) 077301,
  [\href{http://arxiv.org/abs/hep-ph/0601225}{{\tt hep-ph/0601225}}].

\bibitem{Ma:2000cc}
E.~Ma, {\it {Naturally small seesaw neutrino mass with no new physics beyond
  the TeV scale}},  {\em Phys.Rev.Lett.} {\bf 86} (2001) 2502--2504,
  [\href{http://arxiv.org/abs/hep-ph/0011121}{{\tt hep-ph/0011121}}].

\bibitem{Bulbul:2014sua}
E.~Bulbul, M.~Markevitch, A.~Foster, R.~K. Smith, M.~Loewenstein, et~al., {\it
  {Detection of An Unidentified Emission Line in the Stacked X-ray spectrum of
  Galaxy Clusters}},  {\em Astrophys.J.} {\bf 789} (2014) 13,
  [\href{http://arxiv.org/abs/1402.2301}{{\tt arXiv:1402.2301}}].

\bibitem{Boyarsky:2014jta}
A.~Boyarsky, O.~Ruchayskiy, D.~Iakubovskyi, and J.~Franse, {\it {Unidentified
  Line in X-Ray Spectra of the Andromeda Galaxy and Perseus Galaxy Cluster}},
  {\em Phys.Rev.Lett.} {\bf 113} (2014) 251301,
  [\href{http://arxiv.org/abs/1402.4119}{{\tt arXiv:1402.4119}}].

\bibitem{ATLAS:2012gk}
G.~Aad et~al., {\it {Observation of a new particle in the search for the
  Standard Model Higgs boson with the ATLAS detector at the LHC}},  {\em
  Phys.Lett.} {\bf B716} (July, 2012) 1--29,
  [\href{http://arxiv.org/abs/1207.7214}{{\tt arXiv:1207.7214}}].

\bibitem{CMS:2012gu}
S.~Chatrchyan et~al., {\it {Observation of a new boson at a mass of 125 GeV
  with the CMS experiment at the LHC}},  {\em Phys.Lett.B} (July, 2012)
  [\href{http://arxiv.org/abs/1207.7235}{{\tt arXiv:1207.7235}}].

\bibitem{Weinberg:1979sa}
S.~Weinberg, {\it Baryon and lepton nonconserving processes},  {\em Phys. Rev.
  Lett.} {\bf 43} (1979) 1566--1570.

\bibitem{1305.0267v3}
M.~Drewes and J.~U. Kang, {\it The kinematics of cosmic reheating},
  \href{http://arxiv.org/abs/1305.0267}{{\tt arXiv:1305.0267}}. Nuclear Physics
  B 875 (2013) 315-350 and Corrigendum Nucl. Phys. B 875 (2) (2013) 315-350.

\bibitem{Ade:2015xua}
{\bf Planck} Collaboration, P.~A.~R. Ade et~al., {\it {Planck 2015 results.
  XIII. Cosmological parameters}},  \href{http://arxiv.org/abs/1502.01589}{{\tt
  arXiv:1502.01589}}.

\bibitem{Boyarsky:2008xj}
A.~Boyarsky, J.~Lesgourgues, O.~Ruchayskiy, and M.~Viel, {\it {Lyman-alpha
  constraints on warm and on warm-plus-cold dark matter models}},  {\em JCAP}
  {\bf 0905} (2009) 012, [\href{http://arxiv.org/abs/0812.0010}{{\tt
  arXiv:0812.0010}}].

\bibitem{Kolb:1990vq}
E.~W. Kolb and M.~S. Turner, {\it {The Early Universe}},  {\em Front.Phys.}
  {\bf 69} (1990) 1--547.

\bibitem{Hasenkamp:2012ii}
J.~Hasenkamp and J.~Kersten, {\it {Dark radiation from particle decay:
  cosmological constraints and opportunities}},  {\em JCAP} {\bf 1308} (Dec.,
  2013) 024, [\href{http://arxiv.org/abs/1212.4160}{{\tt arXiv:1212.4160}}].

\bibitem{Kusenko:2009up}
A.~Kusenko, {\it {Sterile neutrinos: The Dark side of the light fermions}},
  {\em Phys.Rept.} {\bf 481} (2009) 1--28,
  [\href{http://arxiv.org/abs/0906.2968}{{\tt arXiv:0906.2968}}].

\bibitem{Boyarsky:2014ska}
A.~Boyarsky, J.~Franse, D.~Iakubovskyi, and O.~Ruchayskiy, {\it {Checking the
  dark matter origin of 3.53~keV line with the Milky Way center}},
  \href{http://arxiv.org/abs/1408.2503}{{\tt arXiv:1408.2503}}.

\bibitem{Boyarsky:2014paa}
A.~Boyarsky, J.~Franse, D.~Iakubovskyi, and O.~Ruchayskiy, {\it {Comment on the
  paper "Dark matter searches going bananas: the contribution of Potassium (and
  Chlorine) to the 3.5 keV line" by T. Jeltema and S. Profumo}},
  \href{http://arxiv.org/abs/1408.4388}{{\tt arXiv:1408.4388}}.

\bibitem{Riemer-Sorensen:2014yda}
S.~Riemer-Sorensen, {\it {Questioning a 3.5 keV dark matter emission line}},
  \href{http://arxiv.org/abs/1405.7943}{{\tt arXiv:1405.7943}}.

\bibitem{Jeltema:2014qfa}
T.~E. Jeltema and S.~Profumo, {\it {Discovery of a 3.5 keV line in the Galactic
  Center and a Critical Look at the Origin of the Line Across Astronomical
  Targets}},  {\em Mon.Not.Roy.Astron.Soc.} {\bf 450} (2015) 2143--2152,
  [\href{http://arxiv.org/abs/1408.1699}{{\tt arXiv:1408.1699}}].

\bibitem{Malyshev:2014xqa}
D.~Malyshev, A.~Neronov, and D.~Eckert, {\it {Constraints on 3.55 keV line
  emission from stacked observations of dwarf spheroidal galaxies}},  {\em
  Phys.Rev.} {\bf D90} (2014) 103506,
  [\href{http://arxiv.org/abs/1408.3531}{{\tt arXiv:1408.3531}}].

\bibitem{Canetti:2012kh}
L.~Canetti, M.~Drewes, T.~Frossard, and M.~Shaposhnikov, {\it {Dark Matter,
  Baryogenesis and Neutrino Oscillations from Right Handed Neutrinos}},  {\em
  Phys.Rev.} {\bf D87} (2013), no.~9 093006,
  [\href{http://arxiv.org/abs/1208.4607}{{\tt arXiv:1208.4607}}].

\bibitem{1506.06752v1}
J.~Ghiglieri and M.~Laine, {\it Improved determination of sterile neutrino dark
  matter spectrum},  \href{http://arxiv.org/abs/1506.06752}{{\tt
  arXiv:1506.06752}}.

\bibitem{Drewes:2015eoa}
M.~Drewes and J.~U. Kang, {\it {Sterile neutrino Dark Matter production from
  scalar decay in a thermal bath}},
  \href{http://arxiv.org/abs/1510.05646}{{\tt arXiv:1510.05646}}.

\bibitem{Boyanovsky:2008nc}
D.~Boyanovsky, {\it {Clustering properties of a sterile neutrino dark matter
  candidate}},  {\em Phys.Rev.} {\bf D78} (2008) 103505,
  [\href{http://arxiv.org/abs/0807.0646}{{\tt arXiv:0807.0646}}].

\bibitem{Weinberg:2008xx}
S.~Weinberg, {\em Cosmology}.
\newblock Oxford University Press, 2008.

\end{thebibliography}\endgroup

\end{document}